\documentclass[conference]{IEEEtran}

\IEEEoverridecommandlockouts

\usepackage{color}
\usepackage{graphicx}
\usepackage{amsmath}
\usepackage{amssymb}
\usepackage{algorithm}
\usepackage{algorithmic}
\usepackage{amsmath}
\usepackage{multirow}
\usepackage{booktabs}
\usepackage{array}
\usepackage{amsthm}
\usepackage{stfloats}
\usepackage{caption}
\usepackage{subfigure}
\usepackage{bm}
\usepackage{booktabs}
\usepackage{setspace}
\usepackage{gensymb}
\usepackage{comment}
\usepackage{url}

\newcommand{\be}{\begin{equation}}
\newcommand{\ee}{\end{equation}}
\newcommand{\bea}{\begin{eqnarray}}
\newcommand{\eea}{\end{eqnarray}}
\newcommand{\ba}{\begin{array}}
\newcommand{\ea}{\end{array}}

\renewcommand{\thesubsubsection}{\thesubsection.\arabic{subsubsection}}

\allowdisplaybreaks[4]

\title{Message Passing based Parameter Estimation in Cooperative MIMO-OFDM ISAC Systems
}

\author{\IEEEauthorblockN{Xiaohan Lv$^{\dag}$, Rang Liu$^{\ddag}$, Yi Chen$^{\dag}$, Qian Liu$^{\dag}$, and Ming Li$^{\dag}$}\\
\vspace{-0.0cm}
	\IEEEauthorblockA{$^{\dag}$
		Dalian University of Technology, Dalian, Liaoning 116024, China \\
		E-mail: \texttt{lvxiaohan@mail.dlut.edu.cn, 
			\{chenyi, qianliu, mli\}@dlut.edu.cn }}
	
	\IEEEauthorblockA{$^{\ddag}$
		University of California, Irvine, CA 92697, USA \\ 
		E-mail: \texttt{rangl2@uci.edu }}
        \vspace{-0.0 cm}
}

\def\BibTeX{{\rm B\kern-.05em{\sc i\kern-.025em b}\kern-.08em
    T\kern-.1667em\lower.7ex\hbox{E}\kern-.125emX}}
\begin{document}
\maketitle
\pagestyle{empty}
\thispagestyle{empty}

\begin{abstract} 
In integrated sensing and communication (ISAC) networks, multiple base stations (BSs) collaboratively sense a common target, leveraging diversity from multiple observation perspectives and joint signal processing to enhance sensing performance. This paper introduces a novel message-passing (MP)-based parameter estimation framework for collaborative MIMO-OFDM ISAC systems, which jointly estimates the target’s position and velocity. First, a signal propagation model is established based on geometric relationships, and a factor graph is constructed to represent the unknown parameters. The sum–product algorithm (SPA) is then applied to this factor graph to jointly estimate the multi-dimensional parameter vector. To reduce communication overhead and computational complexity, we employ a hierarchical message-passing scheme with Gaussian approximation. By adopting parameterized message distributions and layered processing, the proposed method significantly reduces both computational complexity and inter-BS communication overhead. Simulation results demonstrate the effectiveness of the proposed MP-based parameter estimation algorithm and highlight the benefits of multi-perspective observations and joint signal processing for cooperative sensing in MIMO-OFDM ISAC systems.
 \end{abstract}


\section{Introduction} 

As wireless technology evolves towards the sixth generation (6G), integrated sensing and communication (ISAC) has emerged as a critical research area \cite{ISAC LR}. Recent interest in collaborative ISAC highlights the advantages of employing networked dual-functional base stations (BSs), which integrate communication and sensing functionalities. Cooperative sensing across multiple BSs leverages multi-perspective spatial diversity and joint signal processing, overcoming limitations such as occlusion, multipath effects, and interference. This collaborative approach significantly enhances the accuracy, range, and reliability of target detection and localization, positioning ISAC as a promising solution for future 6G networks  \cite{ahmediour2022information}.

The authors in \cite{TianyuYang_CMTD} formulate target detection in distributed ISAC networks as a compressed sensing problem, using sparse Bayesian learning (SBL) to detect targets from multiple received signals. Similarly, The authors in \cite{YihanCang_CDMEC} investigate target detection within a mobile edge computing (MEC)-aided ISAC scenario, proposing joint optimization of BSs and sensing devices. However, these studies primarily address binary detection without fully exploiting spatial diversity to enhance range, resolution, and reliability. In \cite{GuangyiLiu_CS}, a refined orthogonal matching pursuit (R-OMP) method combined with joint data processing is proposed to estimate target position and velocity. Nevertheless, these traditional centralized estimation approaches typically incur significant computational complexity and high inter-BS communication overhead, particularly in large-scale cooperative sensing systems.

Motivated by these limitations, this paper proposes a novel cooperative MIMO-OFDM ISAC framework to accurately estimate target parameters. Considering multi-perspective sensing environments, where multiple BSs observe angle-dependent radar cross-sections (RCS) exhibiting random amplitude and phase variations, we design a cooperative ISAC architecture exploiting spatial diversity. Specifically, we first develop a practical joint communication and sensing model tailored to MIMO-OFDM ISAC systems. For sensing, a factor graph representing the propagation geometry is constructed, and the sum-product algorithm (SPA) is employed to jointly estimate multidimensional target parameters under a bilinear signal model. To reduce the high computational and communication costs associated with exact high-dimensional inference, we further introduce a hierarchical message-passing framework employing Gaussian approximations. By utilizing parameterized message distributions and layered computations, the proposed scheme significantly reduces computational complexity and inter-BS communication requirements.  
Simulation results demonstrate the effectiveness of the proposed algorithm and highlight the performance gains achieved through spatial diversity in cooperative MIMO-OFDM ISAC systems.

\begin{figure}[t]
	\centering
	\includegraphics[width= 3.2 in]{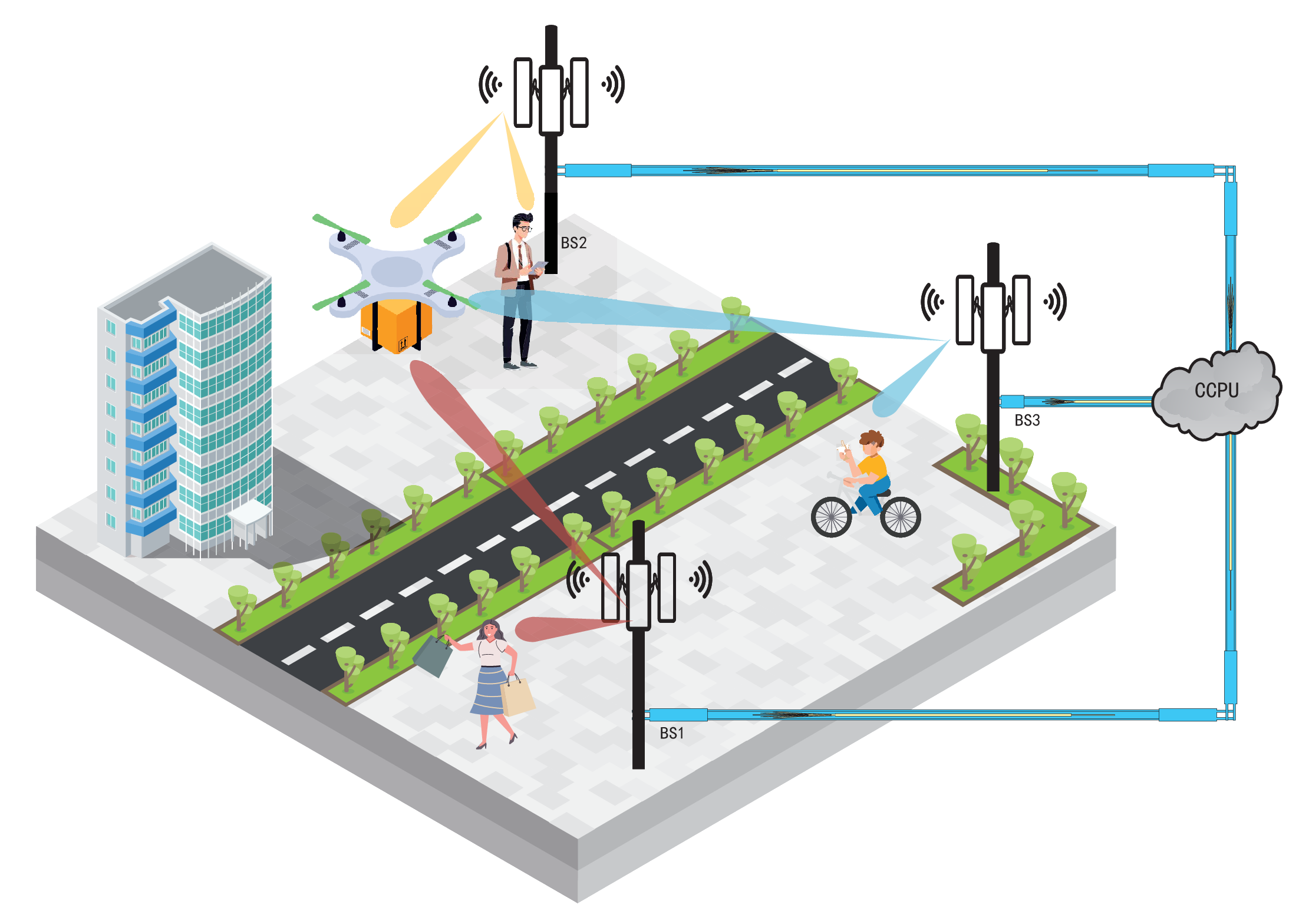}\\
	\caption{A multi-BS collaborative ISAC system.}\label{fig:system model}\vspace{-0.4 cm}
\end{figure} 
\section{System Model and Problem Formulation}
We consider a collaborative MIMO-OFDM ISAC system, as depicted in Fig.~\ref{fig:system model}, which comprises multiple dual-functional BSs. Each BS is equipped with a colocated ULA for both transmission and reception, consisting of 
$N_\text{t}$ transmit and $N_\text{r}$ receive elements, respectively, with half-wavelength spacing between adjacent elements. Each dual-functional BS simultaneously transmits OFDM signals to perform two tasks: communicating with its respective single-antenna user and sensing a common moving target.

\vspace{-0.2 cm}
\subsection{Transmit Signal}
\vspace{-0.1 cm}
The dual-functional baseband signal $\mathbf{x}_{l,n}[m] \in \mathbb{C}^{N_t}$ transmitted by the $l$-th BS on the $n$-th subcarrier of the $m$-th OFDM symbol is expressed as
\begin{equation}\label{precoding process}
\mathbf{x}_{l,n}[m] = \mathbf{w}_{l,n} s_{l,n}[m], ~\forall n, m,
\end{equation}
where $s_{l,n}[m]$ represents the communication symbol for the served user and $\mathbf{w}_{l,n} \in \mathbb{C}^{N_\mathrm{t}}$ represents the transmit beamforming vector.
We assume $M$ OFDM symbols during one coherent processing interval (CPI). By applying an $N$-point inverse fast Fourier transform (IFFT) at the transmitter, the time-domain transmit signal of the $l$-th BS over one CPI is
\begin{equation}\label{transmitted signal}
	{
	\widetilde{\mathbf{x}}_l(t)\triangleq \sum_{n=0}^{N-1} \sum_{m=0}^{M-1} \mathbf{x}_{l,n}[m] e^{\jmath 2\pi n \Delta f t} g_{\mathrm{tx}}(t-mT_{\text{sym}}),}
\end{equation}
where $\Delta f=\frac{1}{T_{\text{u}}}$ represents the subcarrier spacing, $T_{\text{u}}$ denotes the useful OFDM symbol duration, and $T_{\text{sym}}=T_{\text{u}}+T_{\text{cp}}$ is the total symbol duration including cyclic prefix (CP) $T_{\text{cp}}$.
The function $g_{\mathrm{tx}}(t)$ is the rectangular pulse shaping filter. 
The resulting baseband analog signal is up-converted to the radio frequency (RF) domain via $N_\mathrm{t}$ RF chains at carrier frequency $f_{\mathrm{c},l}$ and then emitted through the antennas.

\vspace{-0.3 cm}
\subsection{Radar Echo Signal}
\vspace{-0.1 cm}
For radar sensing function, we consider a scenario where the \(l\)-th BS is located at \((x_l,y_l)\) and a target is positioned at \((x_0,y_0)\) with velocity \((v_x, v_y)\). This work focuses on an open environment without significant multipath effects, where only the direct echo from a single target is considered for parameter estimation. 
Based on the geometric relationships, the round-trip delay \(\tau_l\) and the azimuth angle \(\theta_l\) from the perspective of the \(l\)-th BS can be expressed as
\begin{equation}
	\label{eq:delay}
	\tau_l = \frac{2\sqrt{(x_0-x_l)^2\!+\!(y_0-y_l)^2}}{c},\theta_l = \arctan\left(\frac{y_0-y_l}{x_0-x_l}\right),
\end{equation}
where $c$ represents the speed of light. 
The Doppler shift \(\nu_l\) is given by
\begin{equation}
	\label{eq:doppler}
	\nu_l = 2f_{\mathrm{c},l}/c\left(v_x\cos\theta_l + v_y\sin\theta_l\right),
\end{equation}
where $f_{\mathrm{c},l}$ denotes the carrier frequency of the $l$-th BS.

The transmitted signals propagate to the target and are then reflected back to the receive antenna arrays at each BS. During propagation, the signals experience a round-trip propagation delay \(\tau_l\), a Doppler shift \(\nu_l\), a propagation path loss \(\beta_l\), and a scattering loss \(\alpha_l\) associated with the RCS of the target. The RCS is assumed to be unknown and to exhibit random fluctuations with respect to the observation angle. Consequently, the continuous-time baseband echo signal received at the $l$-th BS is expressed as \cite{OFDM_lps}, \cite{Xiao TSP 2024}
\be	 \label{simplified received baseband echo signal}
\mathbf{y}_l(t)\triangleq \alpha_l\beta_l\mathbf{a}_{\text{r},l}(\theta_l)
\mathbf{a}_{\text{t},l}^H(\theta_l)\widetilde{\mathbf{x}}_l(t-\tau_l )e^{\jmath2\pi \nu_{l}t} + \mathbf{e}(t),
\ee	
where \(\mathbf{e}(t)\) denotes additive white Gaussian noise (AWGN), and the RCS parameter $\alpha_l$ is modeled as a complex Gaussian random variable, i.e., \(\alpha_l\sim\mathcal{CN}(0,\sigma^2_{l})\).
The transmit and receive steering vectors for the antenna arrays are respectively defined as $\mathbf{a}_{\text{t},l}(\theta_l) = [1, e^{-\jmath \pi\sin\theta_l},\dots,e^{-\jmath (N_\text{t}-1)\pi\sin\theta_l}]^T$ and $\mathbf{a}_{\text{r},l}(\theta_l) = [1, e^{-\jmath \pi\sin\theta_l},\dots,e^{-\jmath (N_\text{r}-1)\pi\sin\theta_l}]^T$.

After down-conversion, CP removal, and an 
$N$-point FFT, the frequency-domain baseband echo signal received at the $l$-th BS on the $n$-th subcarrier during the $m$-th OFDM symbol is expressed as
\begin{equation}\label{eq:receive ecch zlnm}
\begin{aligned}
\mathbf{y}_{l,n}[m]&=\alpha_l\beta_l\mathbf{a}_{\text{r},l}(\theta_l)\mathbf{a}_{\text{t},l}^H(\theta_l)\mathbf{x}_{l,n}[m]e^{-\jmath2\pi \tau_l n\Delta f} e^{\jmath2\pi \nu_{l}mT} \\
&  + \mathbf{g}_{l,n}[m],
\end{aligned} \end{equation} 
where $\mathbf{g}_{l,n}[m]\sim\mathcal{CN}(\mathbf{0},\sigma_\text{r}^2\mathbf{I})$ is AWGN.
For notational convenience in subsequent algorithm development, let $y_{l,n,k}[m]$ and $z_{l,n,k}[m]$ denote the noise-corrupted and noise-free observations at the $k$-th antenna, respectively.

To facilitate efficient processing of the radar echoes, we stack the $M$ symbol observations on the $n$-th subcarrier into the matrix
\begin{equation}
\mathbf{Y}_{l,n} \triangleq \big[\mathbf{y}_{l,n}[0],~\mathbf{y}_{l,n}[1],\ldots,\mathbf{y}_{l,n}[M-1]\big]\in\mathbb{C}^{N_r\times M}.
\end{equation}
Substituting $\mathbf{x}_{l,n}[m]=\mathbf{w}_{l,n}s_{l,n}[m]$ into \eqref{eq:receive ecch zlnm} and collecting terms yields the following compact matrix form for the $n$-th subcarrier:
\begin{equation}
\mathbf{Y}_{l,n}
= \alpha_l\beta_l e^{-\jmath2\pi \tau_l n\Delta f}
\mathbf{a}_{\mathrm r,l}(\theta_l)\mathbf{a}_{\mathrm t,l}^{H}(\theta_l)
\mathbf{w}_{l,n}\mathbf{s}_{l,n}^{T}\mathbf{D}_{\nu_l}
+\mathbf{G}_{l,n},
  \end{equation}
  where $\mathbf{s}_{l,n}\triangleq \big[s_{l,n}[0],s_{l,n}[1],\ldots,s_{l,n}[M-1]\big]^{T}$ denotes the transmitted symbol sequence, and
$\mathbf{D}_{\nu_l}\triangleq \mathrm{diag}\{1, e^{\jmath 2 \pi \nu_l T}, \ldots, e^{\jmath 2 \pi \nu_l (M-1)T}\}\in \mathbb{C}^{M\times M}$
  is the Doppler-induced phase rotation matrix. The noise matrix is defined as $\mathbf{G}_{l,n}\triangleq \big[\mathbf{g}_{l,n}[0],\mathbf{g}_{l,n}[1],\ldots,\mathbf{g}_{l,n}[M-1]\big]\in\mathbb{C}^{N_r\times M}$, with
  $\mathrm{vec}(\mathbf{G}_{l,n})\sim \mathcal{CN}(\mathbf{0},\sigma_r^2\mathbf{I}_{N_rM})$.
Next, concatenating $\{\mathbf{Y}_{l,n}\}_{n=0}^{N-1}$ across all $N$ subcarriers, we form the complete received signal matrix at the $l$-th BS as
\begin{equation}
\mathbf{Y}_l \triangleq \big[\mathbf{Y}_{l,0}^{T},~\mathbf{Y}_{l,1}^{T},\ldots,\mathbf{Y}_{l,N-1}^{T}\big]^{T}\in\mathbb{C}^{NN_r\times M}.
\end{equation}
It can be expressed as
\begin{equation}
\mathbf{Y}_l
= \alpha_l\big(\mathbf{I}_N\otimes\mathbf{a}_{\mathrm r,l}(\theta_l)\big)\mathbf{D}_{\tau_l}\mathbf{D}_l\mathbf{S}_l\mathbf{D}_{\nu_l}
+\mathbf{G}_l,
\end{equation}
where $\mathbf{S}_l\triangleq [\mathbf{s}_{l,0}^{T},\ldots,\mathbf{s}_{l,N-1}^{T}]^{T}\in\mathbb{C}^{N\times M}$ denotes the communication symbol matrix, $\mathbf{D}_{\tau_l}$ captures the subcarrier-dependent delay-induced phase shifts, and $\mathbf{D}_l$ collects the subcarrier-dependent transmit spatial gains, given by
\begin{subequations}
\begin{align}
\mathbf{D}_{\tau_l} &\triangleq \mathrm{diag}\{1,e^{-\jmath2\pi \tau_l\Delta f},\ldots,e^{-\jmath2\pi \tau_l(N-1)\Delta f}\},\\
\mathbf{D}_l &\triangleq \beta_l\mathrm{diag}\{\mathbf{a}_{\mathrm t,l}^{H}(\theta_l)\mathbf{w}_{l,0},\ldots,\mathbf{a}_{\mathrm t,l}^{H}(\theta_l)\mathbf{w}_{l,N-1}\}.
\end{align}
\end{subequations}
The overall noise matrix is $\mathbf{G}_l\triangleq [\mathbf{G}_{l,0}^{T},\mathbf{G}_{l,1}^{T},\ldots,\mathbf{G}_{l,N-1}^{T}]^{T}$.

\begin{figure}[!t]
	\centering
	\includegraphics[width= 2.2in]{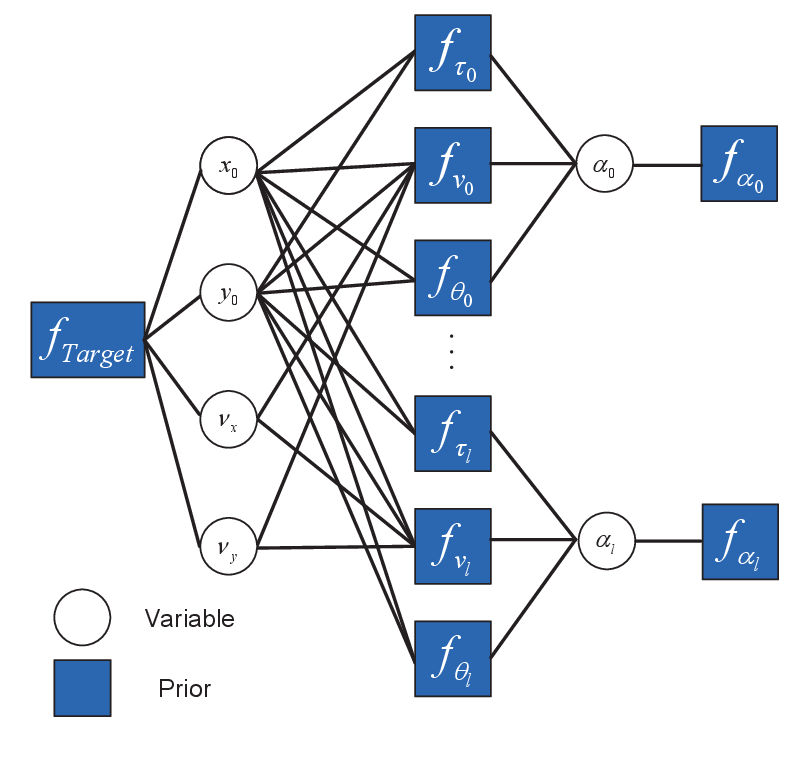}\\
    \vspace{-0.20 cm}
	\caption{Factor graph representation and message passing schedule.}\label{fig:factor graph}\vspace{-0.30 cm}
\end{figure}

\newcounter{TempEqCnt}
\setcounter{TempEqCnt}{\value{equation}}
\setcounter{equation}{16}
\begin{figure*}[!b]
\vspace{-0.30cm}
	\rule[-0pt]{18.1 cm}{0.05em}
	\begin{align}\label{eq:GP}
		&p(\boldsymbol{\xi}, \boldsymbol{\alpha}|\{\mathbf{Y}_l\}_{l=1}^{L}) = \frac{p(\{\mathbf{Y}_l\}_{l=1}^{L}|\boldsymbol{\xi}, \boldsymbol{\alpha})\,p(\boldsymbol{\xi})\,p(\boldsymbol{\alpha})}{p(\{\mathbf{Y}_l\}_{l=1}^{L})} \\[2mm]
        &\propto p(\{\mathbf{Y}_l\}_{l=1}^{L}|\boldsymbol{\xi}, \boldsymbol{\alpha})\,p(\boldsymbol{\xi})\,p(\boldsymbol{\alpha}) \notag 
		= \Big[\prod_{l,n,k,m} p\left(y_{l,n,k}[m]\big|z_{l,n,k}[m]\right)\Big] \Big[\prod_{q=1}^{4}p_{\xi_q}(\xi_q)\Big]\Big[\prod_{l=1}^{L}p_{\alpha_l}(\alpha_l)\Big].
	\end{align}
\end{figure*}
\setcounter{equation}{\value{TempEqCnt}}
\subsection{Problem Formulation}
The goal of this paper is to estimate the target state vector 
\be
\bm{\xi} \triangleq [x_0,y_0,v_x,v_y]^T,
\vspace{-0.20 cm}
\ee
and the complex RCS parameters $\bm{\alpha}\triangleq [\alpha_1,\alpha_2,\ldots,\alpha_L]^T$ based on the received echo signals $\{\mathbf{Y}_l\}_{l=1}^L$ across multiple BSs. Specifically, this problem can be formulated as a maximum a posteriori (MAP) estimation:
\be
\{\hat{\bm{\xi}},\hat{\bm{\alpha}}\} = \arg\underset{\bm{\xi},\bm{\alpha}}{\max}~ p(\bm{\xi},\bm{\alpha}| \{\mathbf{Y}_l\}_{l=1}^L),
\ee
where the posterior distribution, according to Bayes' theorem, can be expanded as
\vspace{-0.30 cm}
\be
p(\bm{\xi},\bm{\alpha}| \{\mathbf{Y}_l\}_{l=1}^L) \propto \big[\prod_{l=1}^L p(\mathbf{Y}_l|\boldsymbol{\xi}, \boldsymbol{\alpha})\big] p(\boldsymbol{\xi}) p(\boldsymbol{\alpha}).\vspace{-0.20 cm}
\ee
Here, the likelihood function $p(\mathbf{Y}_l|\boldsymbol{\xi}, \boldsymbol{\alpha})$ encapsulates the nonlinear and high-dimensional relationships between the received signals and the unknown parameters. The prior distributions $p(\boldsymbol{\xi})$ and $p(\boldsymbol{\alpha})$ represent any available prior knowledge about the target state and RCS parameters, respectively.

The joint estimation of target states and RCS parameters from distributed observations across multiple BSs inherently involves high-dimensional integrations and complex nonlinear couplings. Traditional centralized inference methods suffer from rapidly escalating computational complexity and communication overhead, particularly in large-scale multi-BS scenarios.  
This motivates the development of alternative distributed inference methods, as presented in the next section.

\section{Message Passing for Parameter Estimation}
 To effectively address computational challenges and reduce communication overhead, this section proposes a hierarchical Gaussian-approximated message passing framework. The framework leverages factor graph representations to systematically decompose global inference tasks into efficient local and global message exchanges, significantly reducing complexity and improving scalability.

\subsection{Bayesian Model and Factor Graph Representation}

We begin by formulating the parameter estimation problem within a Bayesian framework. The goal is to estimate the target state parameter vector $\bm{\xi}$ and the local RCS parameter $\alpha_l$ at the $l$-th BS from the observed data $\mathbf{Y}_l$. Assuming independent prior distributions for each component of the target state vector, we have 
\begin{equation}
p(\boldsymbol{\xi}) = \prod_{q=1}^4 p(\xi_q).
\end{equation}
Furthermore, we assume that the conditional likelihood function of the observed signals $\mathbf{Y}_l$, given the noiseless signals $\mathbf{Z}_l$, factorizes across subcarriers, antennas, and symbols, as
\begin{equation}
p(\mathbf{Y}_l|\mathbf{Z}_l) = \prod_{k,n,m} p(y_{l,n,k}[m]|z_{l,n,k}[m]),
\end{equation}
where $z_{l,n,k}[m]$ denotes the noiseless signal received by the $k$-th antenna of the $l$-th BS on the $n$-th subcarrier during the $m$-th symbol.
Consequently, the global posterior distribution can be explicitly expressed using Bayes' theorem as shown in \eqref{eq:GP}, presented at the bottom of this page, which encapsulates the intricate bilinear relationships among the observed signals, target state parameters, and RCS parameters.

These complex statistical relationships represented by the posterior distribution \eqref{eq:GP}, shown at the bottem of  this page, can be effectively visualized and managed through a factor graph representation, as depicted in Fig.~\ref{fig:factor graph}. Specifically, the factor graph comprises two types of nodes:
\begin{itemize}
    \item \textbf{Variable nodes}: Represented by white circles, these nodes correspond to the unknown parameters we aim to estimate, including the components of the target state vector $\bm{\xi}$ and the RCS parameters $\bm{\alpha}$. 
    \item \textbf{Factor nodes}: Represented by blue boxes, these nodes correspond to the factors in the posterior distribution, encompassing the likelihood functions $p(\mathbf{Y}_l|\bm{\xi},\alpha_l)$ and the prior distributions $p(\bm{\xi})$ and $p(\alpha_l)$.
\end{itemize}

In the factor graph, each variable node connects to the factor nodes representing the distributions in which it appears. The observed data $\{y_{l,n,k}[m]\}$ enter the factor nodes as known parameters rather than random variables. This graphical representation decomposes the complicated high-dimensional inference problem into simpler local computations and message exchanges between nodes, naturally enabling distributed inference. 

\setcounter{equation}{17}
\begin{figure*}[b]
\vspace{-0.30cm}
\rule[-0pt]{18.1 cm}{0.05em}
\vspace{-0.50cm}
\begin{subequations}\label{eq:message update}
\begin{align}
\Delta_{l,n,k,m \to q}^{\bm{\xi}}(t, \xi_{q}) &= \log \int_{\alpha_{l},\{\xi_{q'}\}_{q'\neq q}} p\left(y_{l,n,k}[m]\big|z_{l,n,k}[m]\right)\exp\left(\Delta_{l,n,k,m \leftarrow l}^{\alpha_l}(t, \alpha_{l})\right) \notag\\
&\quad\times \prod_{q'\neq q}\exp\left(\Delta_{l,n,k,m \leftarrow q'}^{\bm{\xi}}(t, \xi_{q'})\right)d\alpha_{l} d\{\xi_{q'}\}_{q'\neq q} + \text{const},\\
\Delta_{l,n,k,m \to l}^{\alpha_l}(t, \alpha_{l}) &= \log \int_{\boldsymbol{\xi}} p\left(y_{l,n,k}[m]\big|z_{l,n,k}[m]\right)\prod_{q}\exp\left(\Delta_{l,n,k,m \leftarrow q}^{\bm{\xi}}(t, \xi_{q})\right)d\boldsymbol{\xi} + \text{const},\\
\Delta_{l,n,k,m \leftarrow q}^{\bm{\xi}}(t+1, \xi_{q}) &= \log p(\xi_q) + \sum_{(l',n',k',m')\in\mathcal{F}(q)\backslash(l,n,k,m)} \Delta_{l',n',k',m' \to q}^{\bm{\xi}}(t, \xi_{q}) + \text{const},\\
\Delta_{l,n,k,m \leftarrow l}^{\alpha_l}(t+1, \alpha_{l}) &= \log p(\alpha_l) + \sum_{(n',k',m')\in\mathcal{F}(l)\backslash(n,k,m)}\Delta_{l,n',k',m' \to l}^{\alpha_l}(t, \alpha_{l}) + \text{const}.
\end{align}\vspace{-0.3cm}\end{subequations}
\begin{subequations}\label{eq:pdf update}
\begin{align}
\Delta_{q}^{\bm{\xi}}(t + 1, \xi_{q}) &= \log p(\xi_{q}) + \sum_{(l,n,k,m)\in\mathcal{F}(q)}\Delta_{l,n,k,m \to q}^{\bm{\xi}}(t, \xi_{q}) + \text{const}, \\
\Delta_{l}^{\alpha_l}(t + 1, \alpha_{l}) &= \log p(\alpha_{l}) + \sum_{(n,k,m)\in\mathcal{F}(l)}\Delta_{l,n,k,m \to l}^{\alpha_l}(t, \alpha_{l}) + \text{const}.
\end{align}\end{subequations}

\end{figure*}

\subsection{Sum-Product Algorithm}

SPA is effective for tree-structured factor graphs, providing exact posterior distributions. However, in multi-BS cooperative ISAC scenarios, factor graphs often contain loops due to shared variables like the target state vector $\bm{\xi}$ and RCS parameters $\bm{\alpha}$, causing SPA to approximate rather than compute the exact posterior. Despite lacking guaranteed the theoretical convergence, empirical results show that SPA achieves accurate approximations in sparse or locally tree-like graphs \cite{SPA}.

To improve numerical stability and simplify computation, SPA messages are represented as logarithmic probability density functions (PDF)  with constant offsets. The message $\Delta_{l,n,k,m \to q}^{\bm{\xi}}(t, \cdot)$ at the $t$-th iteration corresponds to the PDF $\frac{1}{C} \exp(\Delta_{l,n,k,m \to q}^{\bm{\xi}}(t,\cdot))$, with normalization factor $C = \int_{\xi_{q}} \exp(\Delta_{l,n,k,m \to q}^{\bm{\xi}}(t,\xi_{q}))$. Four types of SPA messages are summarized in Table~\ref{table:spa_messages}, and target state and RCS distributions at the $t$-th iteration are also represented in the log-domain by $\Delta_{q}^{\bm{\xi}}(t, \cdot)$ and $\Delta_{l}^{\alpha_{l}}(t, \cdot)$.

\begin{table}[t]
\centering
\caption{SPA message definitions at the $t$-th iteration}
\label{table:spa_messages}
\begin{tabular}{|l  l|}
\hline
\(\Delta_{l,n,k,m \to q}^{{\xi}}(t,\cdot)\) &Message from  \(p_{y_{l,n,k}[m]|z_{l,n,k}[m]}\) to \({\xi}_{q}\) \\ \hline 
\(\Delta_{l,n,k,m \leftarrow q}^{{\xi}}(t,\cdot)\) &Message from  \({\xi}_{q}\) to  \(p_{y_{l,n,k}[m]|z_{l,n,k}[m]}\) \\ \hline
\(\Delta_{l,n,k,m \to l}^{{\alpha_l}}(t,\cdot)\) &Message from  \(p_{y_{l,n,k}[m]|z_{l,n,k}[m]}\) to \({\alpha}_{l}\) \\ \hline
\(\Delta_{l,n,k,m \leftarrow l}^{\alpha_l}(t,\cdot)\) &Message from  \(\alpha_{l}\) to  \(p_{y_{l,n,k}[m]|z_{l,n,k}[m]}\) \\ \hline
\(\Delta_{q}^{{\xi}}(t,\cdot)\) &Log posterior PDF of \({\xi}_{q}\) \\ \hline
\(\Delta_{l}^{{\alpha_{l}}}(t,\cdot)\) &Log posterior PDF of \(\mathrm{\alpha}_{l}\) \\ 
\hline
\end{tabular}\vspace{-0.20 cm}
\end{table}

Based on the log-domain message definitions in Table I, the SPA updates at the $t$-th iteration are given in \eqref{eq:message update}. Here, $\mathcal{F}(q)$ denotes the collection of likelihood factor nodes $p\left(y_{l,n,k}[m]\mid z_{l,n,k}[m]\right)$ that involve the state component $\xi_q$, i.e., factors indexed by $(l,n,k,m)$ connected to $\xi_q$. Likewise, $\mathcal{F}(l)$ denotes the set of likelihood factor nodes connected to the RCS variable $\alpha_l$. The set difference $\mathcal{F}(q)\backslash(l,n,k,m)$ and $\mathcal{F}(l)\backslash(n,k,m))$  respectively excludes the specific factor node indexed by $(l,n,k,m)$ and $(n,k,m)$, when computing the corresponding variable-to-factor message. The additive term ``const'' is a normalization constant in the log domain and can be chosen such that the resulting messages (or beliefs) integrate to one.

With these definitions, the log-beliefs (i.e., the log posterior PDFs) of $\xi_q$ and $\alpha_l$ at $t$-th iteration are obtained by combining their priors with all incoming factor-to-variable messages, as summarized in \eqref{eq:pdf update}.

\subsection{Hierarchical Gaussian-Approximated Message Passing}
The global implementation of the SPA described above involves high-dimensional integrals and complex parameter couplings, resulting in prohibitively high communication and computational burden as the number of BSs and subcarriers increases. To address these effectively, we propose a hierarchical message-passing framework enhanced by Gaussian approximations, significantly simplifying the inference process.

Specifically, our hierarchical framework decomposes the global inference task into two distinct stages: local estimation and global aggregation.
At the local stage, the $l$-th BS independently runs a simplified Gaussian-approximated SPA using only its own observations $\mathbf{Y}_l$, and forms the local Gaussian beliefs
$\Delta^{\bm{\xi}}_{l\to q}(t, \xi_q)$ and $\Delta_{l,n,k,m\to l}^{\alpha_l}(t, \alpha_{l})$. 
These parameters  constitute the local messages reported to the fusion center.
At the global stage, a central fusion node aggregates the received local Gaussian beliefs to update the global belief,
thereby producing the refined local state estimate $\Delta_{l,n,k,m \leftarrow q}^{\bm{\xi}}(t+1, \xi_{q})$ and $\Delta_{l,n,k,m \leftarrow l}^{\alpha_l}(t+1, \alpha_{l})$.

\addtolength{\topmargin}{0.05in}

This hierarchical decomposition offers two significant advantages. First, it drastically reduces computational complexity by transforming the originally intractable high-dimensional integrations into straightforward algebraic computations involving only Gaussian parameters. Second, by decoupling local computations from global fusion, it significantly decreases inter-BS communication overhead and enhances parallel processing efficiency. 

Considering that RCS are independent across different irradiation angle, the global posterior distribution at the $l$-th local BS can be expressed as 
\begin{equation}
\begin{aligned}
\!&\!\!\!p(\bm{\xi},\alpha_l | \mathbf{Y}_l) \propto p(\mathbf{Y}_l|\bm{\xi},\alpha_l) p(\bm{\xi}) p(\alpha_l)   \\ 
\!&\!\!\!\!= p(\mathbf{Y}_l|\alpha_{l}(\mathbf{I}_N\!\otimes\!\mathbf{a}_{\text{r},l}(\theta_l))\mathbf{D}_{\tau_l}\mathbf{D}_l\mathbf{S}_l\mathbf{D}_{\nu_l}) \big[ \prod_{q} p(\xi_q) \big] p(\alpha_{l}).\!\!
\end{aligned} 
\vspace{-0.3cm}
\end{equation}
We express the corresponding SPA messages as
\begin{equation}
\begin{aligned}
&\Delta_{l\to q}^{\bm{\xi}}(t, \xi_q) =\log \int_{\alpha_{l},\{\xi_{q'}\}_{q'\neq q}} p(\mathbf{Y}_l|\bm{\xi},\alpha_l)\exp\left(\Delta_{l \leftarrow l}^{\alpha_{l}}(t, \alpha_{l})\right)  \\
 &\qquad \times \prod_{q' \neq q}\exp\left(\Delta_{l \leftarrow q'}^{\bm{\xi}}(t, \xi_{q'})\right) d\alpha_{l}d\{\xi_{q'}\}_{q'\neq q} + \text{const}, \label{eq:mp_s_ltoq} \\
 \end{aligned} 
 \end{equation}
 \begin{equation}
\Delta_{l\leftarrow q}^{\bm{\xi}}(t + 1, \xi_q) = \log p(\xi_q) + \sum_{l=1} \Delta_{l' \to q}^{\bm{\xi}}(t, \xi_q) + \text{const}, 
\end{equation}
where the RCS-related messages are computed and exchanged only in each local BS following update rules in (18d) and (19b).

To further reduce computational complexity, we introduce Gaussian approximated message passing (GAMP) to simplify the SPA message updates \cite{GAMP}. Specifically, we approximate all messages $\Delta^{\bm{\xi}}_{l\to q}(t, \xi_q)$ and $\Delta^{\alpha}_{l}(t, \alpha_{l})$ as Gaussian distributions, parameterized succinctly by their means and variances:
\begin{align}
	&\Delta^{\bm{\xi}}_{l\to q}(t, \xi_q) \propto -\frac{(\xi_q - \mu_{l\to q}^{(t)})^2}{2(\sigma_{l\to q}^{(t)})^2} \\
	&\Delta_{l,n,k,m\to l}^{\alpha_l}(t, \alpha_{l}) \propto -\frac{(\alpha_{l} - \mu_{{l,n,k,m\to l}}^{(t)})^2}{2(\sigma_{{l,n,k,m\to l}}^{(t)})^2},
\end{align}
where $\mu_{l \to q}^{(t)}$ and $(\sigma_{l \to q}^{(t)})^2$ represent the mean and variance estimates of the target state parameter $\xi_q$ at the $l$-th BS, and $\mu_{l,n,k,m \to l}^{(t)}$ and $(\sigma_{l,n,k,m \to l}^{(t)})^2$ are the message parameters associated with the RCS  parameter in the $l$-th base station.

Leveraging Gaussian approximation properties, the integral in (\ref{eq:mp_s_ltoq}) simplifies considerably as \cite{BiG-AMP}
\be
\int_{\alpha_l} \underbrace{\prod_{n,k,m} \exp\left(-\frac{(\alpha_{l} - \tilde{\mu}_{\alpha_{l},n,k,m})^2}{2\tilde{\sigma}_{\alpha_{l},n,k,m}^2}\right)}_{\text{Local Gaussian message for each subcarrier/antenna/symbol}} d\alpha_{l},
\ee
by utilizing two key properties:
\begin{enumerate}
\item \textbf{Gaussian Product Property}: The product of Gaussian distributions remains Gaussian, characterized by new mean and variance values.
\item \textbf{Closed-form Integration}: Integrating Gaussian functions yields closed-form Gaussian results.
\end{enumerate}
Consequently, the high-dimensional integrals in (\ref{eq:mp_s_ltoq}) reduce to algebraic operations involving aggregated Gaussian parameters.

By adopting Gaussian approximations and hierarchical decomposition, the proposed framework converts complex nonlinear integrations into straightforward computations with means and variances. Additionally, local computations are decoupled from global fusion, substantially lowering computational complexity, communication overhead, and enhancing the scalability and practicality of distributed sensing networks.

\section{Simulation Results}
In this section, we present simulation results to evaluate the effectiveness of the proposed collaborative MIMO-OFDM ISAC system and MP-based parameter estimation algorithm. We consider a scenario comprising three BSs located at coordinates $(0,0)$, $(100,100)$, and $(50,100)$ meters. The target position is set to $(60,40)$.  
The other detailed parameters are: $N = 36$, $M=64$, $N_t=N_r = 8$, $P_{l,n} = 5~\text{W}$, $\sigma_c^2 = -20~\text{dB}$, $\sigma_r^2 = -20~\text{dB}$, $\sigma_l^2 = 1$. 

To evaluate the proposed MP-based cooperative localization method (\textbf{``Multi-BS, Proposed, MP''}), we compared its performance with three conventional localization schemes under both single-BS and multi-BS configurations. Specifically, we considered the following benchmark methods:
\begin{itemize} 
\item \textbf{Single-BS}: A single-BS scenario that used maximum-likelihood (ML) for range estimation combined with MUSIC-based angle estimation, representing a near-optimal single-BS benchmark. 
\item \textbf{Multi-BS, Para.-level Fusion}: A multi-BS scenario in which each BS independently estimated the range (via ML) and angle (via MUSIC). The resulting parameter estimates were then fused using a LS method, representing a parameter-level cooperative multi-BS benchmark.
\item \textbf{Multi-BS, Signal-level Fusion}: A multi-BS scenario in which each BS uploaded all received echo signals to a centralized processor for joint position and velocity estimation using ML algorithm, representing an optimal multi-BS benchmark. 
\end{itemize}

\begin{figure}[!t]
\centering
\subfigure[The position estimation RMSE.]{		\includegraphics[width = 1.6 in]{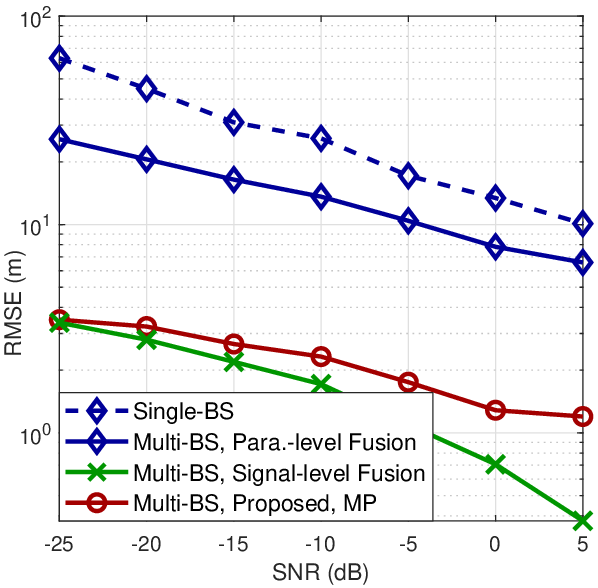}\label{fig:rmse_pos}}
\subfigure[The velocity estimation RMSE.]{
    \includegraphics[width = 1.6 in]{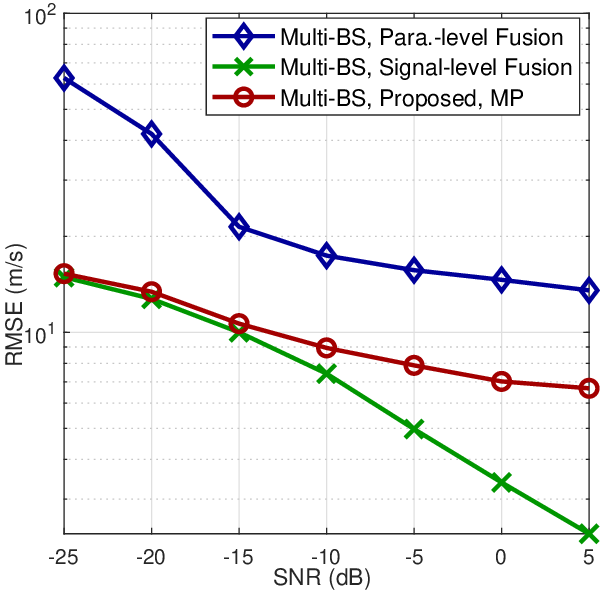}\label{fig:rmse_vel}}
\caption{Sensing and estimation performance versus SNR.}\label{fig:rmse} \vspace{-0.3 cm}
\end{figure}

As shown in Fig.~\ref{fig:rmse}, the proposed MP-based approach consistently achieved low RMSE across a broad range of SNRs, outperforming both the single-BS method and the multi-BS parameter-level fusion strategy. By leveraging multi-perspective sensing (i.e., multiple BSs observing a common target) and iterative refinement, the MP framework effectively mitigated the impact of noise and RCS fluctuations, thereby significantly improving estimation accuracy. Although a centralized signal-level fusion strategy yielded an even lower RMSE, it required collecting the full echo signals from all BSs and thus incurred substantially higher communication overhead and computational complexity.

\begin{figure}[!t]
	\centering
	\includegraphics[width = 3.2 in]{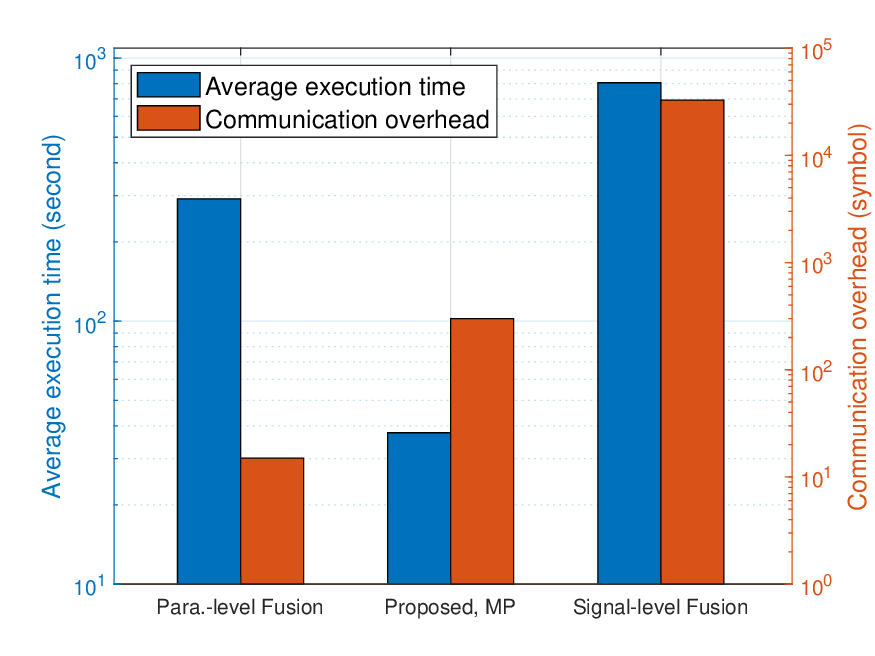}
	\caption{Comparison of computational complexity and communication overhead.}\label{fig:cvso}
    \vspace{-0.3 cm} 
\end{figure}

As illustrated in Fig.~\ref{fig:cvso}, the various multi-BS target estimation approaches were quantitatively compared in terms of communication overhead and computational complexity. The proposed MP-based approach exhibited the lowest computational complexity while maintaining a moderate communication overhead. This efficiency advantage demonstrates the practical effectiveness of the proposed MP-based framework in balancing estimation accuracy with resource usage in cooperative multi-BS localization scenarios.

\begin{figure}[!t]
	\centering
	\includegraphics[width = 3.3 in]{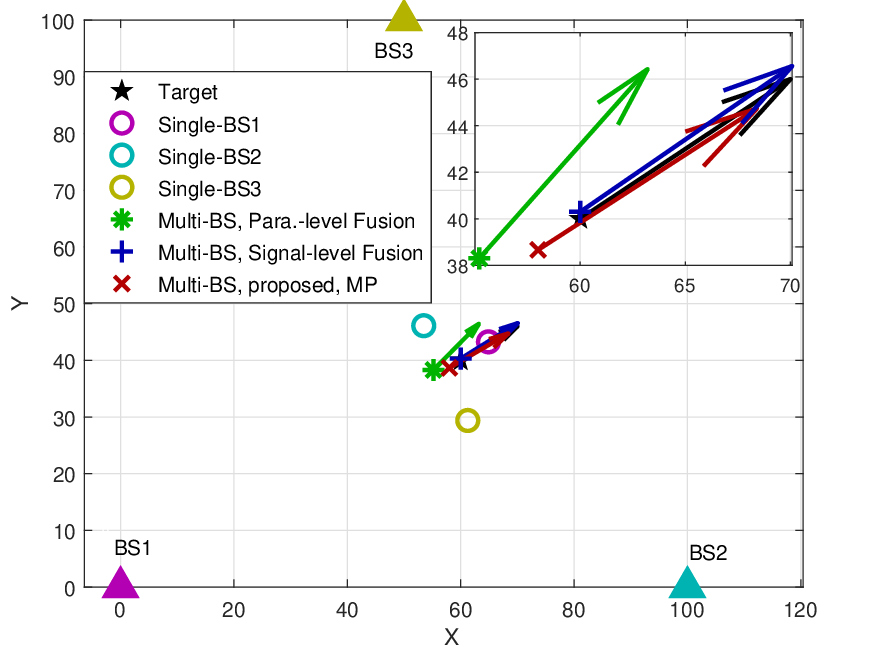}\label{fig:Global_View}
	\caption{Position and velocity estimation results. (Target: $(60, 40)\text{ m}$, $(30, 50)\text{ m/s}$, SNR: $0$ dB).}\label{fig:2D_localization}
\end{figure}

Fig. ~\ref{fig:2D_localization} illustrates the estimation results of the target position and absolute velocity in a single trial.
The markers `$\circ$' represent independent estimation results from individual BSs. The markers `$*$', `$+$', and `$\times$' denote parameter-level fusion, signal-level fusion, and the proposed MP-based algorithm, respectively. It is observed that cooperative sensing significantly outperforms single-BS sensing by providing cooperative gain and overcoming single-view limitations. Moreover, the proposed algorithm achieves performance remarkably close to the signal-level fusion while requiring significantly lower backhaul overhead and computational complexity.

\section{Conclusion}
We investigated high-accuracy target parameter estimation in cooperative MIMO-OFDM ISAC systems. We constructed a signal propagation model based on geometric relationships and designed a factor graph to represent the relationships between the unknown target parameters and the observations. We then employed a message-passing algorithm on this factor graph to jointly estimate the multidimensional target parameters. To address the computational and communication overhead associated with high-dimensional integration, we further proposed a hierarchical message-passing framework based on Gaussian approximation. Simulation results have demonstrated the effectiveness of the proposed algorithm.

\vfill
		
\end{document}